\documentclass[showpacs,prd]{revtex4}
\usepackage{epsfig}
\usepackage{graphicx}
\usepackage{dcolumn}
\usepackage{amsmath}
\usepackage{latexsym}

\begin{document}

\title{Moyal products -- a new perspective on quasi-hermitian quantum mechanics}
\author{F G Scholtz} 
\altaffiliation{fgs@sun.ac.za}
\author{H B Geyer} 
\altaffiliation{hbg@sun.ac.za}
\affiliation{Institute of Theoretical Physics, University of
Stellenbosch,\\ Stellenbosch 7600, South Africa}
\date{\today}

\begin{abstract}
The rationale for introducing non-hermitian Hamiltonians and other observables is reviewed and open issues identified.  We present a new approach based on Moyal products to compute the metric for quasi-hermitian systems.  This approach is not only an efficient method of computation, but also suggests a new perspective on quasi-hermitian quantum mechanics which invites further exploration. In particular, we present some first results which link the Berry connection and curvature to non-perturbative properties and the metric.


\end{abstract}

\maketitle

\section{Introduction}
\label{intro}

The history of quantum mechanics now spans more then a century.  The mathematical underpinning, although more recent, is also well established, particularly due to the work of von Neumann and Weyl, amongst others (see \cite{bratteli} for a comprehensive treatment).  A central result of this analysis, known as the Stone-von Neumann theorem (\cite{bratteli}, vol 2, pp. 34-37), states that for systems with a finite number of degrees of freedom, all unitary irreducible representation of the Weyl algebra (determined by the canonical commutation relations) are equivalent.  In essence this implies that the quantization of systems with a finite number of degrees of freedom is unique, up to unitary transformations. For systems with an infinite number of degrees of freedom the situation is more complex.  In this case there are inequivalent representations of the Weyl algebra and, correspondingly, such systems can be quantized in inequivalent ways.  

Central to the analysis mentioned above, which forms the basis for our current formulation and interpretation of quantum mechanics, is the idea that classical observables should be represented as hermitian operators on the quantum level.  In particular position and momentum are viewed as observables that have to be quantized as hermitian operators.  In the past three to four decades, it has, however, transpired that this may be an unnecessarily restrictive point of view for a variety of reasons.  In particular it turns out that it is convenient to resort to non-hermitian observables, mostly the Hamiltonian, when considering effective interactions \cite{schucan,barrett} or when bosonizing a fermionic system, as the one- plus two-body character of the Hamiltonian can only be maintained in this way \cite{geyer,Doba}, while also maintaining a lowest order association between bosons and fermion pairs.  

More recently, there has been considerable interest in so-called PT-symmetric quantum mechanics in which the Hamiltonian is taken to be non-hermitian, but PT-invariant. Several examples of such Hamiltonians have been found which exhibit a PT-unbroken phase in which the eigenvalues are real and a normal quantum mechanical interpretation is made possible by the introduction of a new inner product, based on the so-called ${\mathcal C}$-operator having properties very similar to the charge conjugation operator \cite{bbj2,bender}.  The possibilty of quantum computing has necessitated the study of open quantum systems in which the Hamiltonian necessarily becomes non-hermitian in order to capture the dissipative nature of the system, see e.g. \cite{qcomp1,qcomp2}. Therefore, the use, and even necessity, of non-hermitian Hamiltonians is nowadays generally accepted, be it for mathematical simplicity or the description of physical reality.

Above we alluded to two quite distinct situations in which non-hermiticity is introduced. In the case of dissipative systems the Hamiltonian will generally have complex eigenvalues and a normal quantum mechanical interpretation is simply not possible, and indeed inappropriate.  In the cases of bosonization, effective interactions and PT-symmetric quantum mechanics, non-hermiticity is introduced with other considerations in mind, such as a simplified mathematical description.  A normal quantum mechanical interpretation should, however, still be possible.  In this article we consider the latter systems, and consider when and how a system of non-hermitian observables constitute a consistent quantum system.  This question is not new and has, in fact, been considered in considerable detail some time ago by the present authors \cite{scholtz}. The conclusion that came out of that analysis was that a system of non-hermitian observables forms a consistent quantum system if a new inner product exists with respect to which the observables are all hermitian. The existence of this new inner product was tied to the existence of a positive definite, bounded hermitian metric operator $\Theta$, defined on the whole Hilbert space, and which obeys the following exchange rule with the observables $A_i$ \cite{scholtz}:
\begin{equation}
\label{metric}
A_i\Theta=\Theta A_i^\dagger\,\,\forall i\,.
\end{equation}
Furthermore it was shown that this metric operator is unique (up to a global normalization) if and only if the set of observables under consideration forms an irreducible set \cite{scholtz}.  All these results apply to the case of a finite number of degrees of freedom.  

The considerations above open up an alternative way of thinking about quantum mechanics. Conventionally one chooses a Hilbert space (and by implication inner product) and the construction of observables is dictated by the requirement of hermiticity with respect to this inner product. With the results above in place, one can, however, take the point of view that the observables are the primary objects which uniquely fix the Hilbert space and inner product.  The latter, of course,  only exists if the observables form a consistent set.  To implement this approach in practice, as is demonstrated in section \ref{examples}, one starts with a Hilbert space which carries an irreducible unitary representation of the Weyl algebra, i.e., hermitian position and momentum operators and thus a unitary representation of the Heisenberg algebra. Then one proceeds to write down the observables of the theory, e.g., energy angular momentum etc. as functions of the Hermitian position and momentum operators, but, in contrast to the conventional approach, the condition that these observables are hermitian w.r.t. the inner product on the Hilbert space is relaxed.  To verify that this set of observables constitute a consistent quantum system one verifies the existence of a metric operator with the properties stated above.  One can also take a slightly different point of view to this construction.  If a metric operator exists, one easily verifies from (\ref{metric}) that under the similarity transformation $\Theta^{-1/2}A_i(\hat x,\hat p)\Theta^{1/2}=A_i(\Theta^{-1/2}\hat x\Theta^{1/2},\Theta^{-1/2}\hat p\Theta^{1/2})$ the observables become Hermitian on the Hilbert space under consideration (see also \cite{kretsch1,kretsch2}).  Thus, one could take the point of view that the observables are hermitian functions of non-hermitian position and momentum operators $\Theta^{-1/2}\hat x\Theta^{1/2}$ and $\Theta^{-1/2}\hat p\Theta^{1/2}$.  From this point of view one thus considers a non-unitary representation of the Weyl algebra, i.e., non-hermitian position and momentum, and construct the observables as hermitian functions of these non-hermitian position and momentum operators. If the metric exists, this non-unitary representation of the Weyl algebra is equivalent to a unitary representation.  This also implies that if the non-unitary representation under consideration is not equivalent to a unitary one, the metric cannot exist.  

Note that since all irreducible unitary representations of the Weyl algebra are equivalent, all non-unitary representations, equivalent to a unitary representation, are also equivalent.  Whether the Weyl algebra admits non-equivalent non-unitary irreducible representations is, to our knowledge, an open question.  However, if such representations were to be encountered in the construction procedure above, the metric cannot exist or the construction procedure has to be revised.  Thus, at least within the construction procedure outlined above, no inequivalent quantizations of a given system (with finite number of degrees of freedom) can occur if it is to constitute a consistent quantum system.   

Given this situation, one may wonder what is to be gained in quantum mechanics from taking the point of view advocated above.  One advantage has already been mentioned, namely, mathematical simplification.  As already pointed out, when bosonising fermionic systems the one-plus-two body character of the Hamiltonian can only be maintained in a non-hermitian description, while the hermitian counterpart contains all possible higher order interaction terms\cite{geyer,Doba}.  Related to this observation, it was recently also noted in the context of PT-symmetric quantum mechanics that the non-hermitian Hamiltonian often lends itself more readily to a perturbative treatment than its hermitian counterpart \cite{bender2}.  This reflects the fact that the implied similarity transformation relating the Hermitian and non-hermitain Hamiltonians already accounts for some non-perturbative effects. These examples illustrate that a judicious choice of observables (which fixes the inner product) may lead to considerable simplification and even the possibility of non-perturbative solutions.      

Central to the succesful practical implementation of the approach outlined above, is the verification of the existence of the metric operator and its subsequent construction.  In this regard it is important to note that if one is only interested in solving the eigenvalue equation for the Hamiltonian, it is sufficient to verify the existence of a metric operator obeying the exchange rule (\ref{metric}) with the Hamiltonian as one is then ensured of the reality of eigenvalues.  For this purpose it is not necessary to construct the metric explicitly as the solution of the eigenvalue equation requires no explicit knowledge of the metric.  However, if one intends to compute other physical quantities such as expectation values, transition probabilities and cross sections, detailed knowledge of the metric is required and the metric must be computed explicitly.  It is also at this point that the uniqueness (up to a global normalization) of the metric becomes important as physical quantities will depend on the choice of metric if it is not unique.  

Solving operator equations of the type (\ref{metric}) is notoriously difficult.  The succesful implementation of the approach outlined above hinges therefore very much on a reliable way of solving these types of equations, or at the very least verifying the existence of a solution with the properties required from a metric operator.  Therefore this issue was partially investigated in \cite{scholtz} and more recently from a variety of different viewpoints in \cite{most1,mosta,bbj,bbj1,jones,swanson}. Here we propose a new approach, based on the Moyal product, which maps the operator equation (\ref{metric}) exactly onto an equivalent differential equation. Generically this equation may be of infinite order, but in many cases of physical interest it turns out to be finite. This equation contains all the information required to construct the metric operator exactly.  In addition, criteria can be formulated to test the hermiticity and positive definitness of the metric directly on the level of this equation, leading to considerable simplification.  

When one considers eq.\ (\ref{metric}) for a reducible set of observables, e.g. the Hamiltonian alone, the metric is not uniquely determined.  However, on the level of the Moyal product formulation the solution of the corresponding differential equation is uniquely determined once an appropriate set of boundary conditions has been specified. On the other hand, as pointed out above, the metric is uniquely determined (up to an irrelevant normalization factor) if eq.\ (\ref{metric}) is considered for an irreducible set of observables.  This suggests an interplay between boundary conditions in phase space and the choice of an irreducible set of physical observables.  This provides us with yet another perspective on the construction of a quantum system, which is the one we explore somewhat more here.  

The paper is organised in the following way.  In section \ref{Moyal} we briefly review the Moyal product with emphasis on the properties we use in later sections. In section \ref{examples} we construct the metric of three non-hermitian PT symmetric Hamiltonians, two of which are exactly solvable.  The issues of uniqueness of the metric, choice of observables and the relation to boundary conditions in phase space are addressed within these models.  In section \ref{conclusions} we conclude with a brief summary of our findings. The present paper extends the results recently published in \cite{schgey}.

\section{The Moyal product}
\label{Moyal}

The Moyal product \cite{moyal} was introduced by Moyal to facilitate Wigner's phase space formulation of quantum mechanics.  More recently it was  revived in the context of non-commutative systems (see e.g. \cite{fairlie}).  In this section we briefly list the essential features of this construction.  A more detailed exposition can be found in \cite{schgey}. In a finite dimensional Hilbert space with dimension $N$ one can construct an unitary irreducible representation of the Heisenberg-Weyl algebra
\begin{equation}
\label{algebra}
gh=e^{i\phi}hg;\quad g^\dagger=g^{-1},\, h^\dagger=h^{-1}\,,
\end{equation}
where $\phi=\frac{2\pi}{N}$.  The operators $U(n,m)=g^nh^m$, with 
$n=0,1,\ldots N-1$ and $m=0,1,\ldots N-1$, form a basis in the space of operators (matrices) on the 
Hilbert space.   Any operator $A$ can therefore be expanded in the form
\begin{equation}
\label{expandf}
A=\sum_{n,m=0}^{N-1} a_{n,m}g^nh^m\,,\quad a_{n,m}=(U(n,m),A)/N\,.
\end{equation}
Making the substitutions $g\rightarrow e^{i\alpha}\,\quad h\rightarrow e^{i\beta}\;,\alpha\,,\beta\in [0,2\pi)$
in the expansion (\ref{expandf}), turns $A$ into a function $A(\alpha,\beta)$, uniquely determined by the 
operator $A$
\begin{equation}
\label{functionf}
A=\sum_{n,m=0}^{N-1} a_{n,m}e^{in\alpha}e^{im\beta}\,.
\end{equation}
An isomorphism with the operator product can be established by defining the Moyal product of functions 
$A(\alpha,\beta)$ and $B(\alpha,\beta)$ \cite{moyal,fairlie}
\begin{equation}
\label{moyal}
A(\alpha,\beta)\ast B(\alpha,\beta)=A(\alpha,\beta)e^{i\phi\stackrel{\leftarrow}{\partial_\beta}
\stackrel{\rightarrow}{\partial_\alpha}}B(\alpha,\beta)\,,
\end{equation}
where the notation $\stackrel{\leftarrow}{\partial}$ and $\stackrel{\rightarrow}{\partial}$ denotes that 
the derivatives act to the left and right, respectively.  On this level operators are replaced by functions, 
as described by (\ref{functionf}), while the non-commutative nature of the operators is captured by the Moyal 
product.  It is easily checked that the Moyal product is associative, as one would expect from the associativity 
of the corresponding operator product.  Once the function $A(\alpha,\beta)$ is given, the coefficients $a_{n,m}$ 
are computed from a simple Fourier transform.  Insertion of these coefficients in (\ref{expandf}) enables the 
reconstruction of the operator.   

Let $A^\dagger(\alpha,\beta)$ and $A(\alpha,\beta)$ denote the functions corresponding to the hermitian conjugate operator $A^\dagger$ and the operator $A$, respectively.  One easily establishes the following relation between these functions \cite{schgey}
\begin{equation}
\label{hc}
A^\dagger(\alpha,\beta)=e^{i\phi\partial_\alpha \partial_\beta}A^\ast(\alpha,\beta)\,.
\end{equation}
From this it follows that an operator is hermitian if and only if
\begin{equation}
\label{ccf}
A^\ast(\alpha,\beta)=e^{-i\phi \partial_\alpha\partial_\beta}A(\alpha,\beta)\,.
\end{equation}

These results are easily generalized to the case of an infinite dimensional quantum system.  In this case a well known irreducible unitary representation of the Heisenberg-Weyl algebra exists \cite{bratteli}
\begin{equation}
\label{weyl}
e^{it\hat p}e^{is\hat x}=e^{i\hbar ts}e^{is\hat x}e^{it\hat p}\,,
\end{equation}
where $\hat x$ and $\hat p$ are the hermitian position and momentum operators satisfying canonical commutation 
relations.  The operators $U(t,s)\equiv e^{it\hat p}e^{is\hat x}$ constitute a complete set \cite{bratteli} and any operator can be expanded as 
\begin{equation}
\label{expandq}
A(\hat x,\hat p)=\int_{-\infty}^{\infty} dsdt\; a(t,s) e^{it\hat p}e^{is\hat x}\,,\quad a(t,s)=\frac{\hbar}{2\pi} 
(U(t,s),A)\,.
\end{equation}
Replacing $\hat x$ and $\hat p$ by real numbers turns $A(\hat x,\hat p)$ into a function $A(x,p)$, 
uniquely determined by $A$
\begin{equation}
\label{functionq}
A(x,p)=\int_{-\infty}^{\infty} dsdt\; a(t,s) e^{itp}e^{isx}\,.
\end{equation}
An isomorphism with the operator product can be established by introducing the Moyal product of functions
\begin{equation}
\label{moyalq}
A(x,p)\ast B(x,p)=A(x,p)e^{i\hbar \stackrel{\leftarrow}{\partial_x}\stackrel{\rightarrow}{\partial_p}}B(x,p)\,.
\end{equation}
On this level we again work with functions, rather than operators, while the non-commutativity of the operators 
is captured by the Moyal product.  As before associativity is easily verified.  Once the function $A(x,p)$ has 
been determined, the function $a(t,s)$ is determined from a Fourier transform.  Insertion into the expansion 
(\ref{expandq}) recovers the operator $A(\hat x,\hat p)$.  

As before, let the functions $A^\dagger(x,p)$ and $A(x,p)$ denote the functions corresponding to the hermitian 
conjugate operator $A^\dagger(\hat x,\hat p)$ and the operator $A(\hat x,\hat p)$, respectively.  One easily establishes the relation \cite{schgey}
\begin{equation}
\label{hcq}
A^\dagger(x,p)=e^{i\hbar \partial_x\partial_p}A^\ast(x,p)\,.
\end{equation}
This implies that an operator is hermitian if and only if
\begin{equation}
\label{ccq}
A^\ast(x,p)=e^{-i\hbar \partial_x\partial_p}A(x,p)\,.
\end{equation}

In what follows we shall often encounter situations where the operator $A$ is a function of $\hat x$, or $\hat p$, 
only.  It is therefore worthwhile to consider this situation briefly.  Consider the case where $A(\hat p)$. 
From (\ref{expandq}) we have
\begin{equation}
\label{ponly}
a(t,s)=\frac{\hbar}{2\pi} (U(t,s),A(\hat p))=\frac{\hbar}{2\pi}tr(e^{-is\hat x}e^{-it\hat p}A(\hat p))
=\delta(s)\int \frac{dp}{2\pi} A(p)e^{-itp}\,.
\end{equation}
Substituting this result in (\ref{functionq}) we note that the function $A(x,p)$ corresponding to the operator 
$A(\hat p)$ is just $A(p)$, i.e., we just replace the momentum operator by a real number. Clearly, the same 
argument applies to $A(\hat x)$.   

An approach related to the one we discuss here was developed in \cite{bender1}, although in that case the 
position and momentum operators are used as a basis to expand the operators.  Compared to the present approach,
the unboundedness of the position 
and momentum operators complicates the proof of completeness. Secondly, the product rule of these 
operators is not as simple as that of the Weyl algebra. This complicates the implementation on the level of 
classical variables.  The current approach therefore seems to be more generic.

\section{Metrics from Moyal products}
\label{examples}
\subsection{Metric equation in Moyal form}

As was already pointed out in the introduction, there are a variety of reasons why the Hamiltonian, and other observables, of a system may be non-hermitian with respect to the inner product on the Hilbert space on which the system is quantized. The central question is then whether a consistent quantum mechanical interpretation remains possible.  This was answered in \cite{scholtz} where it was pointed out that a normal quantum mechanical interpretation is possible 
if a metric operator $\Theta$ exists which has as domain the whole Hilbert space, is hermitian, positive definite and 
bounded, and satisfies the equation
\begin{equation}
\label{metricdef}
H\Theta=\Theta H^\dagger\,,
\end{equation}
where $H$ denotes the Hamiltonian of the system.  Once the existence of such an operator has been established, a new inner product can be defined with respect to which the Hamiltonian is hermitian and a standard quantum mechanical interpretation is possible.  However, as was pointed out in \cite{scholtz}, and also explored in \cite{most1,most2}, the condition (\ref{metricdef}) is not sufficient to fix the metric uniquely, which implies that the quantum mechanical interpretation based on this metric, and the associated inner product, is ambiguous.  The metric is uniquely determined (up to an irrelevant global normalization) if one requires hermiticity of a complete irreducible set of observables, $A_i$ (of which the Hamiltonian may be a member), with respect to the inner product associated with $\Theta$, i.e., it is required that (\ref{metricdef}) holds for all observables \cite{scholtz}:
\begin{equation}
\label{metricdefg}
A_i\Theta=\Theta A_i^\dagger\,\,\forall i\,.
\end{equation}
From this point of view the choice of observables determines the metric and Hilbert space of the quantum 
system uniquely.  Alternatively, one may argue that if a metric, satisfying (\ref{metricdef}), can be found, 
a particular choice of metric determines the allowed set of measurable observables. This is the spirit of 
PT-symmetric quantum mechanics.  

Let us consider the defining equations (\ref{metricdef}) and (\ref{metricdefg}) on the level of the Moyal product formulation.  On this level the observables $A_i$, their hermitian conjugates $A_i^\dagger$ and the metric operator $\Theta$ get replaced by functions $A_i(x,p)$, $A_i^\dagger (x,p)$ and $\Theta(x,p)$ as prescribed in (\ref{functionq}). Note that $A_i(x,p)$ and $A_i^\dagger (x,p)$ are related as in (\ref{hcq}).  In terms of these functions the defining relation (\ref{metricdef}) then reads
\begin{equation}
\label{metricdefm}
H(x,p)\ast \Theta(x,p)=\Theta(x,p)\ast H^\dagger(x,p)\,,
\end{equation}
while (\ref{metricdefg}) reads
\begin{equation}
\label{metricdefgm}
A_i(x,p)\ast \Theta(x,p)=\Theta(x,p)\ast A_i^\dagger(x,p)\,.
\end{equation}        
If the set of observables $A_i$ is irreducible, equation (\ref{metricdefgm}) determines the metric uniquely up to a global normalization factor, without the necessity of additional boundary conditions.  On the other hand, equation (\ref{metricdefm}) does not determine the metric uniquely and it is necessary to impose further boundary conditions to fix the metric up to a possible normalization factor. This demonstrates the already mentioned interplay between the choice of observables and boundary conditions on the metric function $\Theta(x,p)$. To demonstrate this point more clearly we consider the case where we specify $\hat x$ and $\hat p$ as observables.  In this case equation (\ref{metricdefgm}) simply becomes $\Theta^{(1,0)}(x,p)=\Theta^{(0,1)}(x,p)=0$, i.e., the metric is just a constant.  This simply reflects the following facts: (1) that $\hat x$ and $\hat p$ form an irreducible set, which implies that the metric is uniquely determined up to a global normalization factor and (2) that we have chosen $\hat x$ and $\hat p$ hermitian from the outset (insisting on unitary representations of the Heisenberg-Weyl algebra), so that the metric must be proportional to the identity (the original inner product corresponds to $\Theta=1$). 

\subsection{A solvable model}
\label{soluble}

To demonstrate the technique described above, and to illustrate the features discussed thus far, we study in this section a model for which the metric can be computed exactly and analytically.  We proceed by first choosing the Hamiltonian as the only observable, leaving all other possible observables unspecified.  It was already pointed out that this does not fix the metric uniquely and this is demonstrated explicitly below.  Once this has been established, we proceed to augment the Hamiltonian with other choices of observables which, together with the Hamiltonian, form an irreducible set and show that this indeed eliminates the non-uniqueness of the metric.   

The model we study is defined by the following Hamiltonian, written in second quantized form:
\begin{equation}
\label{ham1}
H=\hbar\omega a^\dagger a+\hbar\alpha aa+\hbar\beta a^\dagger a^\dagger\,,
\end{equation}
where $a$ and $a^\dagger$ are bosonic annihilation and creation operators, respectively.  A finite dimensional version of this model was studied in \cite{scholtz} and more recently this model was studied 
in \cite{swanson,geyer1,jones}. This can be rewritten in terms of $\hat x$ and $\hat p$ in the usual way by setting 
$a^\dagger=(\hat x-i\hat p)/\sqrt{2\hbar}$ and $a=(\hat x+i\hat p)/\sqrt{2\hbar}$.  Suppressing an irrelevant real
constant term, which plays no role in the metric equation, this yields
\begin{equation}
\label{ham2}
H=a\hat p^2+b\hat x^2+ic\hat p\hat x\,,\;a=(\omega-\alpha-\beta)/2,\,b=(\omega+\alpha+\beta)/2,\,c
=(\alpha-\beta)\,.
\end{equation}
On the level of the Moyal bracket formulation this becomes
\begin{equation}
\label{clasham1}
H(x,p)=a p^2+b x^2+icpx;\quad H^\dagger(x,p)=a p^2+b x^2-icpx+c\hbar\,.
\end{equation}
Substituting this in (\ref{metricdefm}) and noting that since $H(x,p)$ is polynomial 
the Moyal product truncates, one obtains the following partial differential equation for $\Theta(x,p)$:
\begin{eqnarray}
\label{diffeq1}
&&c\left(\hbar - 2\,i\,p\,x \right) \,\Theta(x,p) \nonumber\\
&&+\hbar\,\left( \left( c\,p - 2\,i\,b\,x \right) \,\Theta^{(0,1)}(x,p)+ 
     \left( c\,x+2\,i \,a\,p\,\right)\Theta^{(1,0)}(x,p) + b\,\hbar\,\Theta^{(0,2)}(x,p)  - 
a\,\hbar\,\Theta^{(2,0)}(x,p) 
\right)=0\, ,
\end{eqnarray}
where the notation $\Theta^{(n,m)}=\partial^{n+m}\Theta /\partial^n x\partial^m p$ has been introduced. 

Note that since no boundary conditions have been specified, the solution is not unique. On the level of the Moyal product the non-uniqueness of the metric therefore resides in the freedom to specify the boundary conditions in (\ref{diffeq1}).  It should, however, be noted that the boundary condition that is to be imposed on (\ref{diffeq1}) can not be arbitrary as the solution has to conform with the conditions of hermiticity and positive definiteness of the 
metric.  Keeping in mind that the metric is uniquely specified once a complete set of observables, hermitian with 
respect to the inner product associated with $\Theta$,  is identified, an interplay between the boundary 
conditions imposed on (\ref{diffeq1}) and the choice of physical observables on the operator level is suggested. In this regard, 
note that a choice of boundary conditions that does not admit a solution conforming to hermiticity and positive 
definitess, constitutes an inconsistent choice of observables, and subsequently an inconsitent quantum system, as 
was pointed out in \cite{scholtz}. On the other hand, if an appropriate choice of boundary conditions, which 
fixes the metric uniquely, is made, both the Hilbert space and allowed observables in the quantum theory are 
uniquely determined.  The allowed observables can indeed be computed by solving (\ref{metricdefgm}) for the 
observables once the metric has been determined.  In this case each choice of boundary condition for solving (\ref{metricdefgm}) corresponds to 
an admissable observable. Of course, it remains to be determined, typically {\it a posteriori}, when a set of such observables, now by construction quasi-hermitian with respect to the same metric which has been uniquely fixed (or chosen) with respect to the Hamiltonian, constitutes an irreducible set together with the Hamiltonian. 

Before proceeding to the detailed solutions of (\ref{diffeq1}), it is useful to comment on some general features of the equation and its solutions. The first issue that we address is the existence of an hermitian metric operator, 
$\Theta$, as required by the definition of the metric operator.  Equation (\ref{diffeq1}) is clearly linear and of the 
form $L\Theta(x,p)=0$, with $L$ a differential operator.  Using $e^{-i\hbar \partial_x\partial_p}x
e^{i\hbar \partial_x\partial_p}=x-i\hbar\partial_p$ and $e^{-i\hbar \partial_x\partial_p}p
e^{i\hbar \partial_x\partial_p}=p-i\hbar\partial_x$, one easily verifies $e^{-i\hbar \partial_x\partial_p}L
e^{i\hbar \partial_x\partial_p}= -L^*$, so that $L^*e^{-i\hbar \partial_x\partial_p}\Theta(x,p)=0$.  On the other 
hand the complex conjugate of (\ref{diffeq1}) reads $L^*\Theta^*(x,p)=0$. Provided that the boundary conditions 
imposed on (\ref{diffeq1}) also satisfy (\ref{ccq}), uniqueness of the solution ((\ref{diffeq1}) is linear) 
implies $\Theta^*(x,p)=e^{-i\hbar \partial_x\partial_p}\Theta(x,p)$.  Thus, provided that the boundary conditions 
imposed are consistent with (\ref{ccq}), the solution of (\ref{diffeq1}), when employed to construct the metric 
operator as described in the previous section, will yield an hermitian metric operator. 

A further property to note is that if $\Theta(x,p)$ is a solution (not necessarily corresponding to an hermitian and positive definite operator) of (\ref{diffeq1}), or equivalently (\ref{metricdefm}), then for arbitrary functions $f(H(x,p))$ and $g(H^\dagger(x,p))$ 
the following is also a solution: $f(H(x,p))\ast \Theta(x,p)\ast g(H^\dagger(x,p))$, where the functions $f$ 
and $g$ are defined through a Taylor expansion involving the Moyal product.  This is most easily checked 
directly on the level of equation (\ref{metricdefm}) using the associativity of the Moyal product and the 
fact that $f(H)\ast H=H\ast f(H)$ and $g(H^\dagger)\ast H^\dagger=H^\dagger\ast g(H^\dagger)$. This is, once again, 
merely a reflection of the non-uniqueness of the solution of (\ref{diffeq1}), which has to be eliminated 
through some appropriate choice of boundary conditions.  As was pointed out earlier the boundary conditions 
cannot be completely arbitrary, but must conform to hermiticity and positive definiteness.  This does not, 
however, eliminate the freedom in (\ref{diffeq1}) completely and more input is required in the form of 
boundary conditions. Indeed, one can easily verify from (\ref{metricdefm}), associativity, and the relation 
(\ref{hcq}), that if $\Theta(x,p)$ is a solution corresponding to an hermitian and positive definite operator, 
so will be $g(H(x,p))\ast \Theta(x,p)\ast e^{i\hbar\partial_x\partial_p}g(H(x,p))^\ast$.
 
Let us now proceed to the detailed solution of (\ref{diffeq1}).  It is not difficult to find an exact solution in the form of the following one parameter family of metrics:
\begin{equation}
\label{onepar}
\Theta(x,p)=e^{r\,p^2 + s\,p\,x + t\,x^2},\,
\end{equation}
where
\begin{equation}
\label{onepar1}
r=\frac{-c \pm {\sqrt{c^2 - 4\,a\,b\,\hbar\,s\left(2i-\hbar s\right)}}}{4\,b\,\hbar}\,,t
=\frac{c \pm {\sqrt{c^2 - 4\,a\,b\,\hbar\,s\left(2i-\hbar s\right)}}}{4\,a\,\hbar}\,,
\end{equation}
$s$ being a free parameter.  Note that the solution has an essential singularity at $\hbar=0$, so that the metric does not exist as a classical object.  Once this solution has been found, a large class of solutions can be constructed as was pointed out earlier.  The non-uniqueness of the solution is, however, already explicit in (\ref{onepar}) and we concentrate on that for the moment.  To see how this freedom can be eliminated through the specification of other observables, consider the situation where we specify the momentum as a further observable. Then, from (\ref{metricdefgm}), $\Theta(x,p)$ also has to satisfy the equation $\Theta^{(1,0)}(x,p)=0$, i.e., it can not depend on $x$. To satisfy this condition we must have $s=0$, $t=0$ and $r=-\frac{c}{2b\hbar}=\frac{\beta-\alpha}{(\omega+\alpha+\beta)\hbar}$, which removes the freedom in (\ref{onepar}).  Furthermore it is clear that the arbitrariness in the solution $g(H(x,p))\ast \Theta(p)\ast e^{i\hbar\partial_x\partial_p}g(H(x,p))^\ast$ is also removed in that the introduction of the function $g$ will lead to an unwanted position dependency of the metric through the position dependency of the Hamiltonian.  Similarly, if one specifies the position as another observable, the metric must be independent of momentum, which is the case when $s=0$, $r=0$ and $t=\frac{c}{2b\hbar}=\frac{\alpha-\beta}{(\omega+\alpha+\beta)\hbar}$.  As before the introduction of the function $g$ introduces an unwanted momemtum dependency in the metric, so that all freedom, apart from a global normalization factor, has been eliminated.  Note that both these solutions also have the correct asymptotic behaviour in the limit $c\rightarrow 0$.  As a final remark we point out that the first of these solutions would have resulted by imposing the boundary conditions that on a line $p=p_0$ the metric is a constant, while the second results from imposing the same condition on a line $x=x_0$.

Let us now consider the hermiticity and positive definiteness of these solutions. Since $a$, $b$ and $c$ are real, the condition (\ref{ccq}) is trivially satisfied in both these cases and the metric is hermitian. To show positive definiteness one has to verify that the logarithm of the metric is hermitian. In the Moyal product formulation this implies that one has to find the function corresponding to the logarithm of the metric operator and verify that it satisfies (\ref{ccq}), i.e., one has to find the function $\eta(x,p)$ such that $1+\eta+\frac{1}{2!}\eta\ast\eta+\frac{1}{3!}\eta\ast\eta\ast\eta\ast+\ldots=\Theta$.  In this case it is, however, obvious that the Moyal product reduces to an ordinary product so that the function corresponding to the logarithm of the metric operator is simply the logarithm of (\ref{onepar}), which is $-\,c\,p^2/2b\hbar$ or $\,c\,x^2/2a\hbar$, depending on the choice of additional observables $p$ or $x$.  This trivially satisfies (\ref{ccq}) so that the metric is positive definite, although not necessarily bounded.             

As a final example of how the non-uniqueness is removed through the specification of other observables, consider the case where the number operator $\hat N=a^\dagger a=\frac{1}{2\hbar}(\hat p^2+\hat x^2)$ is specified as the other observable. On the level of the Moyal product $\hat N$ gets replaced by $N=\frac{1}{2\hbar}(p^2+x^2)$.  From (\ref{metricdefgm}) and (\ref{clasham1}) we then note that $\Theta(x,p)$ has to satisfy an additional equation, which is identical to (\ref{diffeq1}) with the specific choice of parameters $a=b=\frac{1}{2\hbar}$ and $c=0$.  The solution of this equation is clearly also of the form (\ref{onepar}) where $r=t=\pm i\sqrt{\hbar s(2i-\hbar s)}/2\hbar$.  Since the metric has to satisfy both (\ref{diffeq1}) and this equation, the value of the free parameter $s$ is fixed to be $s_{\pm}=\frac{1}{\hbar}(i\pm\frac{\sqrt{c^2-(a-b)^2}}{|a-b|})=\frac{i}{\hbar}(1\pm\frac{2\sqrt{\alpha\beta}}{|\alpha+\beta|})$ and $r=t=\frac{c}{2\hbar(a-b)}=-\frac{(\alpha-\beta)}{2\hbar(\alpha+\beta)}$.  Only the $s_-$ solution yields the correct asymptotic behaviour for $\Theta(x,p)$ in the hermitian limit $c\rightarrow 0$ or $\alpha\rightarrow\beta$. One can also easily verify that no further freedom exists by introducing the function $g$.  The reason is simply that these functions can only depend on either $H$ or $N$ if one wants a solution of either of these two equations.  Since both equations have to be satisfied this freedom is thus eliminated. 

Finally we check the hermiticity and positive definiteness of this metric, i.e., we check whether the metric and its logarithm satisfy (\ref{ccq}).  We first check the metric.  Since it is an exponential it is very difficult to verify the hermiticity directly.  Instead, the way we do it is to check the hermiticity order by order in a series expansion in $c$. As an example we show the calculation to third order in $c$.  To this order we have for $\Theta(x,p)$ 
\begin{eqnarray}
\label{3rd}
\Theta(x,p)&=&1 + \frac{c\,\left( p^2 + x^2 \right) }{2\hbar(a - b)} + 
  \frac{c^2\,\left( p^4 + 4\,i \,\hbar\,p\,x + 2\,p^2\,x^2 + x^4 \right) }{8\,{\left( a - b \right) }^2\,\hbar^2} + 
  \frac{c^3\,\left( p^2 + x^2 \right) \,\left( p^4 + 12\,i \,\hbar\,p\,x + 2\,p^2\,x^2 + x^4 \right) }
   {48\,{\left( a - b \right) }^3\,\hbar^3}\nonumber\\
   &\equiv& 1+ca_1+c^2a_2+c^3a_3\,.
\end{eqnarray}   
It is now simple to verify that (\ref{ccq}) is indeed satisfied to this order in $c$, verifying the hermiticity of the metric. Next we turn to the logarithm of the metric.  To compute the logarithm to third order we simply expand the logaritm of (\ref{3rd}), keeping in mind that the expansion must be done through the Moyal product in order to reflect the operator nature correctly.  One easily finds
\begin{equation}
\label{logT1}
(\log \Theta)(x,p)=\log(1+ca_1+c^2a_2+c^3a_3)=a_1c+c^2\left(a_2-\frac{a_1\ast a_1}{2}\right)+c^3\left(a_3+\frac{_1\ast a_1\ast a_1}{3}-
\frac{a_1\ast a_2+a_2\ast a_1}{2}\right)\,.
\end{equation}       
Substituting from (\ref{3rd}) the values for $a_1,a_2$ and $a_3$, the logarithm of the metric can easily be evaluated to this order to be 
\begin{equation}
\label{logTsol}
(\log \Theta)(x,p)=\frac{c^2}{4(a-b)^2}+\left(\frac{c}{2\hbar(a-b)}\left(1+\frac{c^2}{3(a-b)^2}\right)\right) \left(p^2+x^2\right)\,.
\end{equation}  
It is now simple to verify that (\ref{ccq}) is satisfied by (\ref{logTsol}), verifying the positive definiteness of the metric.  We note that (\ref{logTsol}) has a simple linear form in $N=p^2+x^2$. It is not too difficult to see that this is indeed a feature that persists to all orders so that in this model the logaritm of the metric has the simple form $(\log \Theta)(x,p)=a(c)+b(c)N$ where $a(c)$ and $b(c)$ are real and $a(0)=b(0)=0$.  The condition (\ref{ccq}) can then in fact be verified to all orders in $c$.  This confirms the positive definiteness and hermiticity to all orders.

\subsection{A shifted oscillator: the $ix$ potential}

Not unexpectedly an exact solution can be obtained for the metric $\Theta$ (and equivalently for the $\mathcal C$-operator) when the Hamiltonian is the PT-symmetric complex shifted harmonic oscillator with $H= \tfrac{1}{2}p^2 + \tfrac{1}{2}x^2 + ix$. In this case it has been shown \cite{bbj} that the general form of the $\mathcal C$-operator,  ${\mathcal C} = e^Q{\mathcal P}$, simplifies to ${\mathcal C} = e^{-2p}{\mathcal P}$. It is well known that the metric can be related to a similarity transformation that transforms a given non-hermitian Hamiltonian into and equivalent hermitian form (see also section \ref{intro}), and that the similarity transformation is also related to the $\mathcal C$-operator) \cite{most1,mosta,scholtz}, which allows the identification  $\Theta = e^{-2p}$ in the present example.

This result also immediateley follows from the general analysis in section \ref{Moyal}. In particular, eq.\ (\ref{metricdefm}) reduces to the partial differential equation
\begin{equation}
\label{linpot}
2ix\Theta (x,p) + (ix-1)\Theta^{(0,1)} - \tfrac{1}{2}\Theta^{(0,2)} - i\Theta^{(1,0)}p + \tfrac{1}{2} \Theta^{(2,0)} = 0,
\end{equation}
where we have set $\hbar = 1$. Assuming $\Theta (x,p) = \Theta (p)$ only, eq.\ (\ref{linpot}) becomes the ordinary differential equation (primes indicating differentiation with respect to $p$)
\begin{equation}
2ix\Theta + (ix-1)\Theta^{\prime} - \tfrac{1}{2}\Theta^{\prime\prime}  = 0,
\end{equation}
with the solution $\Theta (p) = e^{-2p}$, as before.

As already pointed out, the non-uniqueness of the metric determined from the star product (\ref{metricdefm}) is associated with different boundary conditions. For the PDE (\ref{linpot}) above, another class of solutions may accordingly be investigated by assuming  $\Theta (x,p) = \Theta (x)$. The differential equation reduces to a Schr\"odinger equation with a linear potential, with solutions the standard Airy functions, here with complex argument. These solutions, to what extent they conform to the requirements of Hermiticity and positive definiteness, and their physical implications, will be discussed elsewhere \cite{geysch}.

\subsection{The $ix^3$ potential}
\label{PT}   
Next we consider the PT-symmetric model with Hamiltonian \cite{bender}
\begin{equation}
\label{ham}
H(\hat x,\hat p)=\hat p^2 +ig\hat x^3\,.
\end{equation}
Here we consider the Hamiltonian as the only observable and leave the other possible observables 
unspecified.  We then proceed to investigate the existence of a (non-unique) hermitian and positive definite metric. Since the PT-symmetry is unbroken for this Hamiltonian \cite{bender} we expect such a metric to exist.  

On the level of the Moyal product this Hamiltonian and its hermitian conjugate get replaced by the functions (it is a sum of functions depending on 
$\hat p$ and $\hat x$ only)
\begin{equation}
\label{hamf}
H(x,p)=p^2 +igx^3\,, H^\dagger(x,p)=H^\ast(x,p)\,,
\end{equation}
while the metric becomes a function $\Theta(x,p)$ as defined in (\ref{functionq}).  Note that in this case the hermitian conjugate of the operator gets replaced by the complex conjugate of the function $H(x,p)$.  It is simple to see that this is a generic feature of functions (or the sum of functions) that depend on $\hat x$ or $\hat p$ only, as there is no phase due to the exchange of the operators $e^{it\hat p}$ and $e^{is\hat x}$ (see (\ref{ponly})). 

As before the Moyal product in equation (\ref{metricdefm}) truncates due to the polynomial nature of $H(x,p)$ to yield the following differential equation for $\Theta(x,p)$:
\begin{equation}
\label{diffeq}
2\,i \,g\,x^3\,\Theta(x,p) - 3\,g\,\hbar\,x^2\,\Theta^{(0,1)}(x,p) - 
  3\,i \,g\,\hbar^2\,x\,\Theta^{(0,2)}(x,p) + g\,\hbar^3\,\Theta^{(0,3)}(x,p) - 
  2\,i \,\hbar\,p\,\Theta^{(1,0)}(x,p) + \hbar^2\,\Theta^{(2,0)}(x,p)=0\,.
\end{equation}
   
Solutions with the correct asymptotic behaviour can be constructed by resorting to a series representation in 
$g$ for the solutions of (\ref{diffeq}).  This also allows us to make contact with existing literature in 
which a series expansion was used \cite{most2}.  In the same spirit as the solvable model this series expansion can be tested for hermiticity and positive definiteness by using the criterion (\ref{ccq}).

We list the result to $O(g)$ below, as the expression becomes rather elaborate at higher order, but 
in principle it is quite simple to compute the result to any order desired:
\begin{eqnarray}
\label{solog}
&&\Theta(x,p)=\frac{-21\,i \,e^{\frac{2\,i \,p\,x}{\hbar}}\,g\,\hbar^4\,c1(p)}{16p^6} - 
  \frac{i \,e^{\frac{2\,i \,p\,x}{\hbar}}\,\hbar\,c1(p)}{2p} - 
  \frac{21\,e^{\frac{ 2\,i  \,p\,x}{\hbar}}\,g\,\hbar^3\,x\,c1(p)}{8\,p^5} + 
  \frac{i\,e^{\frac{2\,i \,p\,x}{\hbar}}\,g\,\hbar^2\,x^2\,c1(p)}{8p^4} + \nonumber\\
 && \frac{e^{\frac{ 2\,i  \,p\,x}{\hbar}}\,g\,\hbar\,x^3\,c1(p)}{4\,p^3} + c2(p) + 
  \frac{3\,i \,g\,\hbar^2\,x\,c2(p)}{4p^4} - \frac{3\,g\,\hbar\,x^2\,c2(p)}{4\,p^3} - 
  \frac{i\,g\,x^3\,c2(p)}{2p^2} + \frac{g\,x^4\,c2(p)}{4\,\hbar\,p} - 
  \frac{i\,e^{\frac{2\,i \,p\,x}{\hbar}}\,g\,\hbar\,c3(p)}{2p} + \nonumber\\
  && g\,c4(p) + \frac{21\, i\,e^{\frac{2\,i  \,p\,x}{\hbar}}\,g\,\hbar^4\,
     c1'(p)}{16p^5} + \frac{21\,e^{\frac{ 2\,i  \,p\,x}{\hbar}}\,g\,\hbar^3\,x\,c1'(p)}
   {8\,p^4} - \frac{9\,i\,e^{\frac{2\,i  \,p\,x}{\hbar}}\,g\,\hbar^2\,x^2\,c1'(p)}
   {8p^3} - \frac{e^{\frac{2\,i \,p\,x}{\hbar}}\,g\,\hbar\,x^3\,c1'(p)}{4\,p^2} - \nonumber\\
  &&\frac{3\,i\,g\,\hbar^2\,x\,c2'(p)}{4p^3} + \frac{3\,g\,\hbar\,x^2\,c2'(p)}{4\,p^2} + 
  \frac{i\,g\,x^3\,c2'(p)}{2p} - 
  \frac{9\,i \,e^{\frac{2\,i \,p\,x}{\hbar}}\,g\,\hbar^4\,c1''(p)}{16p^4} - 
  \frac{9\,e^{\frac{2\,i \,p\,x}{\hbar}}\,g\,\hbar^3\,x\,c1''(p)}{8\,p^3} + \nonumber\\
  &&\frac{\,i\,e^{\frac{2\,i \,p\,x}{\hbar}}\,g\,\hbar^2\,x^2\,c1''(p)}{8p^2} + 
 \frac{3\,i \,g\,\hbar^2\,x\,c2''(p)}{4p^2} - \frac{3\,g\,\hbar\,x^2\,c2''(p)}{4\,p} + 
 \frac{i \,e^{\frac{2\,i  \,p\,x}{\hbar}}\,g\,\hbar^4\,c1^{(3)}(p)}{8p^3} + \nonumber\\
  &&\frac{e^{\frac{ 2\,i  \,p\,x}{\hbar}}\,g\,\hbar^3\,x\,c1^{(3)}(p)}{4\,p^2} - 
  \frac{i\,g\,\hbar^2\,x\,c2^{(3)}(p)}{2p}\,+O(g^2)\,.
\end{eqnarray}
This is the most general form of the solution where $c1(p)$, $c2(p)$, $c3(p)$ and $c4(p)$ are completely 
arbitrary functions of $p$. These functions are, however, restricted if one requires that the expansion (\ref{solog}) 
satisfies the hermiticity condition (\ref{ccq}).  Indeed, one can quite easily see that $c2(p)$ and $c4(p)$ 
must at least be real.  
Here we do not pursue the most general solution, but rather focus on a particular choice of integration 
constants for which this expression simplifies considerably. Setting $c1(p)=0$, $c2(p)=1$ (this brings about 
only a global normalization), $c3(p)=0$ and $c4(p)=0$ one finds to $O(g)$
\begin{equation}
\label{solbcg}
\Theta(x,p)=1 + g\,\left( \frac{3\,i\,\hbar^2\,x}{4p^4} - \frac{3\,\hbar\,x^2}{4\,p^3} - \frac{i \,x^3}{2p^2} + 
     \frac{x^4}{4\,\hbar\,p} \right)\,.
\end{equation}
This result agrees with the result in \cite{most2} when the different normalization of the kinetic energy  term, 
accounting for the factor of $\frac{1}{2}$, and the different convention for the definitions of the metric 
(see (\ref{metricdef})), which brings about a complex conjugation, are taken into account.  With this choice of 
boundary conditions, even the higher order term simplifies considerably and one easily finds to $O(g^3)$ 
(setting any further integration constants zero)   

\begin{eqnarray}
\label{solbcg2}
\Theta(x,p)&=&1 + g\,\left( \frac{3\,i \,\hbar^2\,x}{4p^4} - \frac{3\,\hbar\,x^2}{4\,p^3} - \frac{i\,x^3}{2p^2} + 
     \frac{x^4}{4\,\hbar\,p} \right) \nonumber\\
     & +& g^2\,\left( \frac{108\,i  \,\hbar^5\,x}{p^9} - \frac{108\,\hbar^4\,x^2}{p^8} - 
     \frac{ 57\,i  \,\hbar^3\,x^3}{p^7} + \frac{21\,\hbar^2\,x^4}{p^6} + 
     \frac{6\,i  \,\hbar\,x^5}{p^5} - \frac{11\,x^6}{8\,p^4} - \frac{i \,x^7}{4\hbar\,p^3} + 
     \frac{x^8}{32\,\hbar^2\,p^2} \right)  \nonumber\\
     &+& g^3\,\left( \frac{29872557\,i \,\hbar^8\,x}{256p^{14}} - 
     \frac{29872557\,\hbar^7\,x^2}{256\,p^{13}} - \frac{7676559\,i\,\hbar^6\,x^3}{128p^{12}} + 
     \frac{5395599\,\hbar^5\,x^4}{256\,p^{11}} + \frac{727299\,i\,\hbar^4\,x^5}{128p^{10}}\right.\nonumber\\
     & -& \left.\frac{159489\,\hbar^3\,x^6}{128\,p^9}
     -\frac{14679\,i \,\hbar^2\,x^7}{64p^8} + \frac{9207\,\hbar\,x^8}{256\,p^7} + 
     \frac{615\,i\,x^9}{128p^6} - \frac{343\,x^{10}}{640\,\hbar\,p^5} - 
     \frac{3\,i \,x^{11}}{64\hbar^2\,p^4} + \frac{x^{12}}{384\,\hbar^3\,p^3} \right)\,+O(g^4)\nonumber\\
     &\equiv& 1+ga+g^2b+g^3c+O(g^4)\,.
\end{eqnarray}   
Note, as in the solvable model, that $\Theta(x,p)$ is singular in the limit $\hbar\rightarrow 0$. 

We can now proceed to check the hermiticity of $\Theta$ by verifying that (\ref{ccq}) holds for (\ref{solbcg2}).  It is easily verified that this is indeed the case up to $O(g^3)$.  

Finally, we consider the positive definiteness of $\Theta$.  For this we have to show that the logarithm of 
$\Theta$ is also hermitian.  As in the solvable model the logarithm of $\Theta$ can be computed from (\ref{solbcg2}) by using (\ref{logT1}). This yields
\begin{eqnarray}
\label{logT2}
(\log \Theta)(x,p)&=&g\,\left( \frac{3\,i\,\hbar^2\,x}{4p^4} - \frac{3\,\hbar\,x^2}{4\,p^3} - \frac{i \,x^3}{2p^2} + 
     \frac{x^4}{4\,\hbar\,p} \right)\nonumber\\
     &+&g^3\left(\frac{-2745171\,i\,\hbar^8\,x}{256p^{14}} + \frac{2745171\,\hbar^7\,x^2}{256\,p^{13}} + 
  \frac{677457\,i\,\hbar^6\,x^3}{128p^{12}} - \frac{439857\,\hbar^5\,x^4}{256\,p^{11}} - 
  \frac{52029\,i\,\hbar^4\,x^5}{128p^{10}} \right.\nonumber\\
  &+&\left. \frac{9375\,\hbar^3\,x^6}{128\,p^9} + 
  \frac{651\,i\,\hbar^2\,x^7}{64p^8} - \frac{273\,\hbar\,x^8}{256\,p^7} - \frac{5\,i \,x^9}{64p^6} + 
  \frac{x^{10}}{320\,\hbar\,p^5}\right)\,,
\end{eqnarray}
which can easily be checked to satisfy (\ref{ccq}). Note that the second order term vanishes with this choice of boundary conditions \cite{most2}.  We have now verified the hermiticity and positive definiteness of the solution, at least to third order in the coupling $g$.

\section{The Berry connection and curvature}

In this section we show how the formalism developed in section \ref{Moyal} can be used to construct the Berry connection and curvature \cite{berry,simon,shapwilcz}.  Although we cannot verify directly from this the existence of a metric operator for a given Hamiltonian, we can identify points or lines in parameter space where singularities will occur in the metric, eliminating the possible existence of a metric in these cases.  This enables us to make statements about the analytic properties of the eigenvalues and eigenstates, which is a non-perturbative piece of information providing information on the radius of convergence of a perturbative treatment.

Let us start by setting up the equation for the Berry connection.  Consider a Hamiltonian depending on a set of real parameters $q_1,q_2\ldots q_n$, denoted $H(q)$. Here we are interested in non-hermitian Hamiltonians and therefore the Hamiltonian does not have to be hermitian as we range over the parameter space.  Furthermore we do not even insist on real eigenvalues of the Hamiltonian as we range over parameter space, in which case a metric can of course not exist.  We now write the Hamiltonian as
\begin{equation}
\label{diag}
H(q)=S(q)D(q)S^{-1}(q)\,,  
\end{equation}
where $D(q)$ is a diagonal operator. It is important to note here that the operator $S(q)$ may be singular at certain points in parameter space. These are the points where the Hamiltonian is not diagonalizable (it can at best be brought in the Jordan form) and the eigenstates do not span the whole space, i.e., they are linearly dependent.  At these points the metric can also not exist (if it does, one can construct an operator that diagonalizes the Hamiltonian).  To proceed we differentiate (\ref{diag}) with respect to $q_i$
\begin{equation}
\label{diff}
\frac{\partial H(q)}{\partial q_i}-S(q)\frac{\partial D(q)}{\partial q_i}S^{-1}(q) =[A_i(q),H(q)]\,.
\end{equation}   
Here we have introduced the Berry connection defined by
\begin{equation}
\label{berryc1}
A_i(q)=\frac{\partial S(q)}{\partial q_i}S^{-1}(q)\,.
\end{equation}
The Berry connection simply generates the change in the eigenstates as the parameters on which the Hamiltonian depend are changed. Furthermore one expects singularities in $S(q)$ to show up in the Berry connection.  Thus solving for the Berry connection from (\ref{berryc1}) one may be able to identify these singular points.  Note that as the components of the Berry connection are operators they do not necessarily commute.    
  
Equation (\ref{diff}) is not very useful in its given form, as we need to know the eigenvalues in order to compute the Berry connection from it.  To avoid this, we take the commutator of (\ref{diff}) with the Hamiltonian.  It is simple to see that $[S(q)\frac{\partial D(q)}{\partial q_i}S^{-1}(q),H(q)]=S(q)( D(q)\frac{\partial D(q)}{\partial q_i}-\frac{\partial D(q)}{\partial q_i}D(q))S^{-1}(q)=0$ and we have
\begin{equation}
\label{diff1}
[\frac{\partial H(q)}{\partial q_i},H(q)] =[[A_i(q),H(q)],H(q)]\,.
\end{equation}   
We can now use this equation to solve for the Berry connection.  Once again this is an operator equation which is generally difficult to solve.  On the level of the Moyal product this becomes, however, again a differential equation of finite order if the Hamiltonian is polynomial.

Before attempting to solve (\ref{diff1}) in a particular model, we first have to consider some of its general features.  It is clear that (\ref{diff1}) does not have a unique solution as we can always add a solution of the homogenous equation to find another solution.  This freedom is indeed already present in equation (\ref{diag}) as a 'gauge freedom'.  To see this we note that if $S(q)$ diagonalizes $H(q)$, so will $T(q)=S(q)\Lambda(q)$ where $\Lambda(q)$ is an arbitrary diagonal operator.  Under the transformation $S(q)\rightarrow T(q)$, the Berry connection transforms to
\begin{equation}
\label{trans}
A_i(q)\rightarrow A_i^\prime (q)=A_i(q)+S(q)\frac{\partial\Lambda(q)}{\partial q_i}\Lambda^{-1}(q)S^{-1}(q)\,,
\end{equation}
illustrating the non-uniqueness of the Berry connection.  Quantities like the Berry phase should, however, be unique and thus invariant under the transformation (\ref{trans}).  Let us therefore consider this issue in some more detail.  To do this we consider the change in $S(q)$ as we translate around a small plaquette in the $q_i$ and $q_j$ directions as shown in figure \ref{plaq}. 
\setlength{\unitlength}{1mm}
\begin{figure}
\begin{picture}(53,53)
\put(-30, 0){\epsfig{file=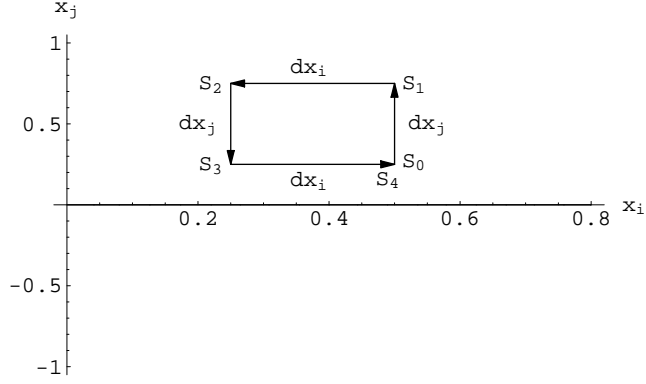, height=53mm}}
\end{picture}
\caption {Change in $S(q)$  with a translation around a plaquette in the $q_i$ and $q_j$ directions.  The initial value of $S(q)$ is denoted by $S_0$ and the final value by $S_4$.  Intermediate values are denoted $S_1$, $S_2$, and $S_3$, repectively.}
\label{plaq}
\end{figure}
Expanding to second order in $dq_i$ and $dq_j$, one easily finds the following relations between the intermediate values (see figure \ref{plaq}) of $S(q)$
\begin{eqnarray}
\label{transl}
S_1&=&\left(1+A_jdq_j+\frac{1}{2}\left(\frac{\partial A_j}{\partial q_j}+A_j^2\right)dq_j^2\right)S_0\,,\nonumber\\
S_2&=&\left(1+A_idq_i+\frac{\partial A_i}{\partial q_j}dq_i\,dq_j+\frac{1}{2}\left(\frac{\partial A_i}{\partial q_i}+A_i^2\right)dq_i^2\right)S_1\,,\nonumber\\
S_3&=&\left(1-A_jdq_j-\frac{\partial A_j}{\partial q_i}dq_i\,dq_j+\frac{1}{2}\left(A_j^2-\frac{\partial A_j}{\partial q_j}\right)dq_j^2\right)S_2\,,\nonumber\\
S_4&=&\left(1-A_idq_i+\frac{1}{2}\left(A_i^2-\frac{\partial A_i}{\partial q_i}\right)dq_i^2\right)S_3\,.\nonumber\\
\end{eqnarray}
Successive application now yields the relation between $S_4$ and $S_0$
\begin{equation}
\label{berrycur1}
S_4=\left(1+F_{ij}dq_idq_j\right)S_0\,,
\end{equation}
where we have introduced the Berry curvature
\begin{equation}
F_{ij}=\frac{\partial A_i}{\partial q_j}-\frac{\partial A_j}{\partial q_i}+[A_i,A_j]\,.
\end{equation}
This is the infinitesimal change in the eigenstates.  For finite closed loops the change can be obtained from a path ordered exponential of the Berry connection or a careful exponentiation of (\ref{berrycur1}), keeping in mind that the $F_{ij}$ at different points need not commute.  It is now a simple matter to verify that under the transformation (\ref{trans}) the Berry curvature is invariant, i.e., $F_{ij}^\prime=F_{ij}$.

Before applying (\ref{diff1}) to infinite dimensional quantum systems, we consider a simple 2-dimensional matrix problem. Consider the following Hamiltonian
\begin{equation}
H(q_1,q_2)=\left( {\begin{array}{*{20}c}
   1 & {q_1  + iq_2 }  \\
   {q_1  + iq_2 } & { - 1}  \\
\end{array} } \right)
\end{equation} 
One can easily solve (\ref{diff1}) to find the most general solution 
\begin{eqnarray}
A_1(q_1,q_2)&=&\left( {\begin{array}{*{20}c}
   \frac{2\,{w_1}}{{q_1} + i \,{q_2}} - 
     \frac{1}{\left( {q_1} + i \,{q_2} \right) \,
        \left( 1 + \left(q_1 + i q_2\right)^2 \right)} + {y_1} & {w_1} - \frac{1}{ 1 + \left(q_1 + i q_2\right)^2}  \\
   w_1 & y_1  \\  
\end{array} } \right)\nonumber\\
A_2(q_1,q_2)&=&\left( {\begin{array}{*{20}c}
   \frac{ 2\,i  \,{w_1}}{{q_1} + i \,{q_2}} - 
    \frac{i }{\left( {q_1} + i \,{q_2} \right) \,
       \left( 1 + \left(q_1 + i q_2\right)^2 \right) } + {y_2} &  i \,{w_1} - \frac{i }{ 1 + \left(q_1 + i q_2\right)^2 } \\
   iw_1& y_2  \\
\end{array} } \right)
\end{eqnarray}
The homogenous part has been constructed to yield a zero curvature.  We note that this expression has singularities at the points $\{q1=0,\,q2=0\}$ and $\{q1=0,\,q2=\pm 1\}$.  The singularity at the origin is spurious (the transformation $S(q)=1$ at this point) and can be removed by an appropriate choice of the homogenous part.  Indeed choosing $w_1=1/2$ and $y_1=y_2=0$ we have 
\begin{eqnarray}
\label{2dim}
A_1(q_1,q_2)=-iA_2(q_1,q_2)=\left( {\begin{array}{*{20}c}
    \frac{{q_1} + i \,{q_2}}{ 1 + \left(q_1 + i q_2\right)^2 }   &  \frac{1}{2}-\frac{1}{1 + \left(q_1 + iq_2\right)^2} \\
   \frac{1}{2} & 0  \\  
\end{array} } \right)
\end{eqnarray}
The singularities at the points  $\{q_1=0,\,q_2=\pm 1\}$ are, however, not removable and, not surprisingly, are the two exceptional points where the matrix is not diagonalizable \cite{kato} and the transformation $S(q)$ becomes singular.  When one computes the curvature it vanishes everywhere, except at these points where the curvature has to be computed with great care due to the singularity.  This implies that moving in an infinitesimal loop around any point brings the eigenfunctions back to themselves; the only exception being the exceptional points where the eigenfunctions transform into each other \cite{kato}.  Note that the Berry connection at different points does not commute.  Due to this fact one has to compute a path ordered exponential in order to obtain the change of the eigenvectors as one moves around in a closed loop. The result (\ref{2dim}) can also be obtained from (\ref{berryc1}) by direct diagonalization and an appropriate choice of 'gauge'.  

To demonstrate the aforementioned points, we compute the curvature at the exceptional point $\{q_1=0,q_2=1\}$.  We do this by computing the change of the eigenstate when we move in an infinitesimal circle around the exceptional point. Setting $q_1=r\cos\phi$ and $q_2=1+r\sin\phi$, we easily compute the Berry connection in the azimuthal direction from (\ref{berryc1}) and (\ref{2dim}) to be $A_\phi=ire^{i\phi} A_1$. This result is now expanded to lowest order in $r$ to obtain
\begin{equation}
\label{berryex}
\frac{\partial S(\phi)}{\partial \phi}S^{-1}(\phi)|_{(0,1)}=A_\phi=\left( {\begin{array}{*{20}c}
   \frac{i}{2}  & -\frac{1}{2}  \\
   0& 0  \\  
\end{array} } \right) \, .
\end{equation}
The change in the eigenstate can now easily be computed from (\ref{berryex}) by noting that $S(2\pi)=(1+\frac{2\pi}{N}A_\phi)^NS(0)$ in the limit $N\rightarrow\infty$.  The result is $S(2\pi)=FS(0)$ with 
\begin{equation}
F=\left( {\begin{array}{*{20}c}
   -1 & -2i  \\
   0& 1  \\  
\end{array} } \right)\,.
\end{equation}
We now proceed to compute the action of $F$ on the eigenstates infinitesimally close to the exceptional point.  These eigenstates, expanded to leading order in $r$, read
\begin{equation}
u_\pm=\left( {\begin{array}{*{20}c}
   { - i \pm i\sqrt {2w} }  \\
   0  \\
\end{array} } \right)\,,
\end{equation}
where we have set $w=re^{i\phi}$.  It is now a simple matter to see that $Fu_\pm=u_\mp$. 

Let us now consider the quantum system discussed in section \ref{soluble}.  All the considerations of section \ref{Moyal} apply and on the level of the Moyal formulation we can replace equation (\ref{diff1}) by 
\begin{equation}
\label{diff1m}
\left[\frac{\partial H(x,p,q)}{\partial q_i},H(x,p,q)\right]_\ast =[[A_i(x,p,q),H(x,p,q)],H(x,p,q)]_\ast\,.
\end{equation}    
Here $x$ and $p$ are the phase space variables, $q$ the parameters in the Hamiltonian and we have introduced the Moyal bracket
\begin{equation}
[A(x,p),B(x,p)]_\ast=A(x,p)\ast B(x,p)-B(x,p)\ast A(x,p)\,.
\end{equation}

For the system of section \ref{soluble} the Hamiltonian was given in (\ref{clasham1}) and the two parameters entering the Hamiltonian are the parameters $\alpha$ and $\beta$ (see (\ref{ham1}) and (\ref{ham2})).  As we are not interested in a global rescaling of the Hamiltonian, which only effects the homogenous part of the Berry connection, it is more convenient to change to two new variables, $q_1$ and $q_2$, by dividing the Hamiltonian (\ref{clasham1}) by $a$:
\begin{equation}
\label{clasham2}
H(x,p)=p^2+q_1 x^2+iq_2 px;\quad q_1=\frac{b}{a}=\frac{\omega+\alpha+\beta}{\omega-\alpha-\beta}\,,q_2=\frac{c}{a}=\frac{2(\alpha-\beta)}{\omega-\alpha-\beta}.
\end{equation}

It is now easy to set up the explicit equations for the connection from (\ref{diff1m}), but we do not list them here due to their length.  We rather make an assumption for the connection of the following form
\begin{equation}
A_i(x,p)=r_ip^2+s_ixp+t_ix^2\,;i=1,2\,,
\end{equation}
and solve for the parameters $r_i$, $s_i$ and $t_i$ as functions of $q_1$ and $q_2$.  The result is
\begin{eqnarray}
A_1(x,p)&=&\frac{i \,p\,x}
   {\hbar\,\left( 4\,{q_1} + {{q_2}}^2 \right) } - 
  \frac{x^2\,{q_2}}
   {2\,\hbar\,\left( 4\,{q_1} + {{q_2}}^2 \right) }\,,\nonumber\\
A_2(x,p)&=&\frac{x^2\,{q_1}}
   {\hbar\,\left( 4\,{q_1} + {{q_2}}^2 \right) } + 
  \frac{i\,p\,x\,{q_2}}
   {2\hbar\,\left( 4\,{q_1} + {{q_2}}^2 \right) }\,.
\end{eqnarray}  
The curvature can also be easily computed from $F_{1,2}=\frac{\partial A_1}{\partial q_2}-\frac{\partial A_2}{\partial q_1}+[A_1,A_2]_\ast$. It vanishes everywhere, except at possible singularities.  Let us therefore investigate the singularities and interpret them.  There are singularities on the curves $4q_1+q_2^2=\omega^2-4\alpha\beta=0$. It is well known \cite{geyer1} that on these curves the Bogoliubov transformation that diagonalizes the Hamiltonian becomes singular, which is now simply reflected on the level of the Berry connection. These continuous curves divide the $\alpha$-$\beta$ parameter space into disjoint regions and it is not possible to move from the one to the other without crossing a singularity of $S(q)$.  This signals that perturbation theory is limited to within these regions and that the radius of convergence will be set by the distance to the closest point on these curves.  This is illustrated in figure \ref{fig2}.  

One may also evaluate these results from the perspective of the existence of the metric $\Theta$. While $S(q)$ is singularity free, except on the border between shaded and unshaded areas, $\Theta$ does not necessarily exist elsewhere, since positive definiteness and hermiticity of $\Theta$ cannot be immediately inferred. At the same time the singularity of $S(q)$ on the borders indicate the existence of zero norm states, implying that on these borders the metric definitely does not exist, confirming in this particular case what was anticipated in general at the beginning of this section.

\setlength{\unitlength}{1mm}
\begin{figure}
\begin{picture}(53,53)
\put(-10, 0){\epsfig{file=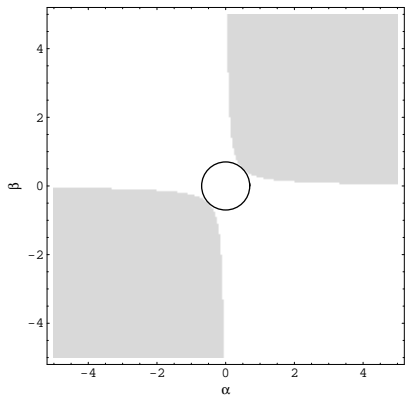, height=53mm}}
\end{picture}
\caption {Singular curves of the Berry connection.  These curves divide the parameter space in disjoint regions, shown as shaded and unshaded here, and it is not possible to pass from one region to another without crossing a singularity in $S(q)$. The circle denotes the expected radius of convergence of a perturbative expansion around the origin.}
\label{fig2}
\end{figure}

\section{Conclusions}
\label{conclusions}
We have presented a brief overview of considerations pertinent tot non- and quasi-hermitian quantum mechanics in which the role of the metric is stressed. Subsequently we have shown how the Moyal product can be used to compute the metric for a given non-hermitian Hamiltonian from a general partial differential equation. The verification that the metric is hermitian and positive definite can be carried out directly on the level 
of the Moyal product formulation, without refering to the operator level. 

We have carried through this program 
for three Hamiltonians, all of which possess PT-symmetry.  These considerations can also be applied to finite 
dimensional models as those studied in \cite{scholtz}, by using the finite dimensional formulation of the Moyal 
product  where, essentially, $h\rightarrow\frac{1}{N}$.  An interesting new perspective that arises from 
the present formulation is the relation between the choice of observables and boundary conditions imposed on the 
metric equation, as formulated in terms of the Moyal product, both of which give rise to a unique metric. We have also dicussed the Berry connection in the context of the Moyal product and demonstrated that non-perturbative information can be extracted from the resulting equation.

\section*{Acknowledgements}

The authors acknowledge financial support from the National Research Foundation of South Africa. We thank Dieter Heiss for useful discussions.

\end{document}